\documentclass[twoside,fleqn,espcrc2]{article}
\usepackage{graphicx,epsfig,pstricks,fancybox,fancyhdr,epic,rotating,color}
\usepackage{amsfonts}

\begin{document}

\begin{center}
{\color{blue} \mbox{ }}
\vspace*{0.0cm}

{\color{blue} \large \bf 
Addenda 2008 to variationsQCD2007 (2nd)
}
\vspace*{0.2cm}

{\color{red} \bf
On concise hypotheses for the interpretation of a wide scalar resonance
as gauge boson binary in QCD $\rightarrow$ some new analyses
}
\vspace*{0.2cm}

\color{blue}
{\small \bf Peter Minkowski}
\\
{\small \bf University of Bern}
\\
\vspace{0.0cm}

{\color{green} \small \bf 
after QCD2008, Montpellier, 6.-13. July 2008 
%\hspace*{0.2cm} $\rightarrow$}
}

\end{center}

%\newpage

{\color{blue} 

\vspace*{-0.2cm}
%\hspace*{4.0cm}
\begin{center}
\hspace*{0.0cm}
\begin{figure}[htb]
% \label{fig1}
\vskip -0.2cm
\hskip -2.0cm
\includegraphics[angle=0,width=5.5cm]{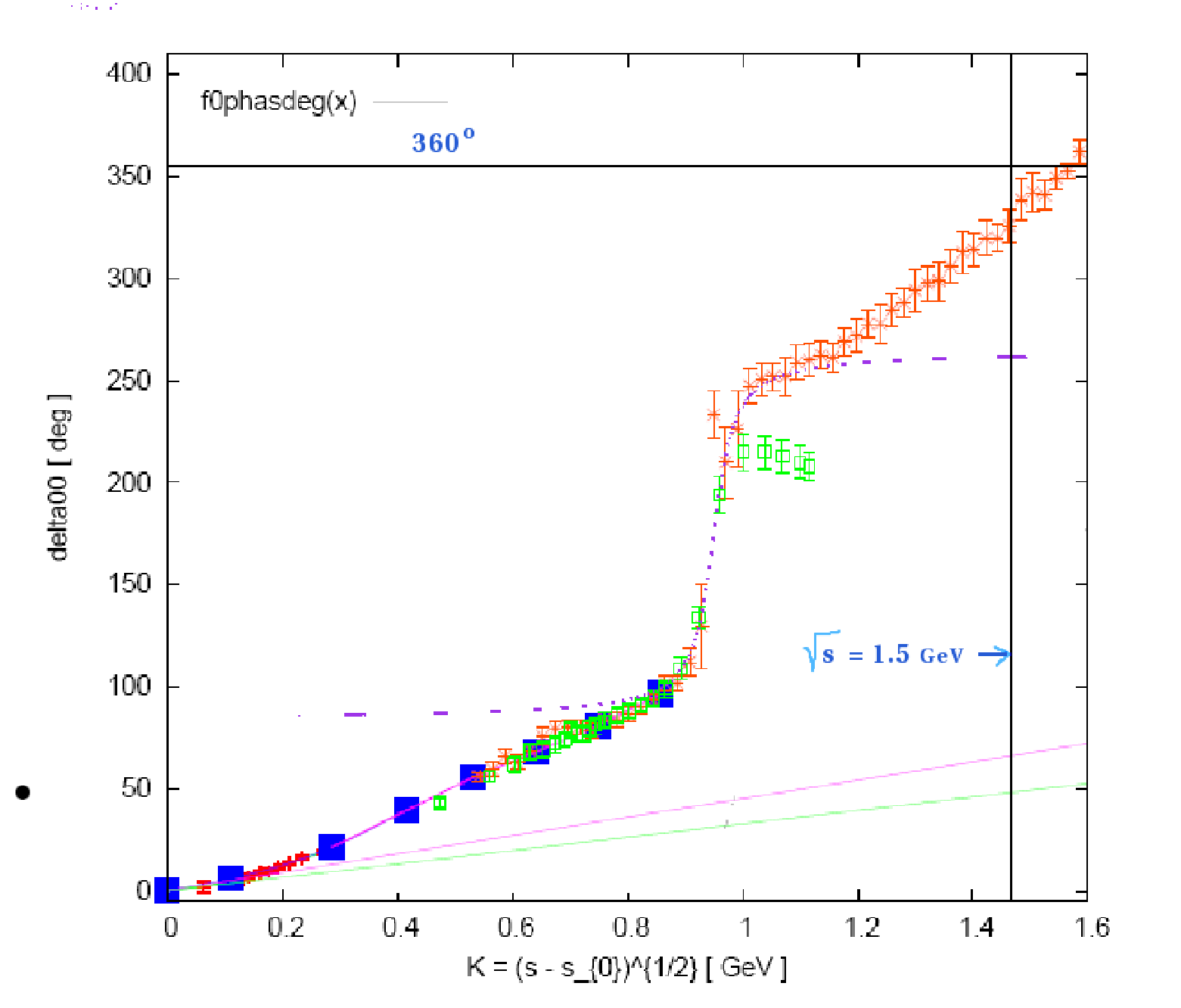}
\vskip -0.2cm
%\hskip -2.0cm
%\caption{}
%\label{fig15a}
\vspace*{+0.20cm}
{\color{red} \hspace*{-2.0cm}
\begin{tabular}{l} Fig. 1 :
\color{red} The $\pi \pi \ , \ I=0 $ {\it ideal} elastic  
 \vspace*{-0.1cm} \\
\color{red}
s-wave from threshold to $\ \sim \ 1.625 \ \mbox{GeV}$.
\end{tabular}
}
\end{figure}
\end{center}
\vspace*{-6.3cm}  \begin{tabular}[t]{@{\hspace*{5.5cm}}l}
This work aims to give some answers to questions 
\vspace*{-0.0cm} \\
raised at QCD2008 \cite{PMQCD2008} . Fig. 1 plots the s-wave
\vspace*{-0.0cm} \\
phase shifts versus 
$K \ = \ \left ( M_{\ \pi \pi}^{\ 2} \ - \ 4 \ m_{\ \pi}^{\ 2} 
 \right )^{ 1/2}$.
\vspace*{-0.0cm} \\
\rule{1mm}{1mm} : from ref. \cite{CGL} {\it Colangelo, Gasser and Leutwyler} ,
\vspace*{-0.0cm} \\
{\color{magenta} \rule{1.5mm}{0.2mm}} \hspace*{0.1cm} interpolates 
\hspace*{0.0cm} \rule{1mm}{1mm} ,
{\color{red} +} : from ref. \cite{Brigitte}
{\it Na48/2 coll.} 
\vspace*{-0.0cm} \\
corrected for isospin breaking ,
{\color{green} \rule[1.5mm]{1.5mm}{0.2mm} \hspace*{-4.2mm}
\rule[-0.1mm]{1.9mm}{0.2mm} \hspace*{-3.2mm} \rule{0.2mm}{1.5mm}
} : from ref. \cite{Protopop73} 
\vspace*{-0.0cm} \\
{\it Protopopescu et al.} , {\color{red} \rule[1.5mm]{1.5mm}{0.2mm}
\hspace*{-3.9mm}
\rule[-0.1mm]{1.5mm}{0.2mm} \hspace*{-3.2mm} \rule{0.2mm}{1.5mm}
} : from ref. \cite{hyams73} {\it CERN-}
\vspace*{-0.0cm} \\
{\it -Munich coll. ; W. Ochs , thesis 1973} .
\vspace*{-0.0cm} \\
{\color{magenta} \rule{1.00mm}{0.2mm} \rule{1.00mm}{0.2mm}
\hspace*{0.2mm} \rule[2mm]{1.00mm}{0.2mm} \rule[2mm]{1.00mm}{0.2mm} 
} : minimal meromorphic parametrization 
\vspace*{-0.0cm} \\
of the influence of f0(980) \hspace*{0.9cm} {\color{magenta} $\rightarrow$}
\vspace*{-0.0cm} \\
{\color{magenta} \rule[1.0mm]{2mm}{0.1mm}} ,
{\color{green} \rule[1.0mm]{2mm}{0.1mm}} linear approximations
$\ \delta00 \ = \ 0.5 \ a00 \ K \ $ ,
\vspace*{-0.0cm} \\
$\leftrightarrow$
\hspace*{0.2cm} {\color{magenta} $a00 \ m_{\ \pi} \ = \ 0.22$} , 
\hspace*{0.2cm} {\color{green} $a00 \ m_{\ \pi} \ = \ 0.16$} .
\vspace*{-0.0cm} \\
{\color{red} {\it ideal}} \hspace*{0.1cm} in the caption to figure 1 refers to
\vspace*{-0.0cm} \\
the limit $\ e \ = \ 0 \ , \ m_{\ d} \ = \ m_{\ u} \ $ . 
\end{tabular}

\rule[-0.1cm]{11.5cm}{0.2mm}
\vspace*{0.1cm}

\hspace*{-2.6cm}
The rapid phase variation induced by f0(980) defines two fringes,
denoted low and high, \\
\hspace*{-2.0cm} the two regions 

\begin{equation}
\label{eq:1}
\hspace*{-2.7cm}
\begin{array}{lll ll} 
\mbox{low} \ : & 0 \ \leq \ K \ \leq \ \sim \ 0.9 \ \mbox{GeV}
& ; &
\mbox{high} \ :  & \sim \ 1.0 \ \mbox{GeV} \ \leq \ K \ \leq \ \sim \ 1.6
\ \mbox{GeV}
\vspace*{0.1cm} \\
&  2 m_{\ \pi} \ \leq \ \sqrt{s} \ \leq \ \sim \ 0.94 \ \mbox{GeV}
& ; & & \sim \ 1.04 \ \mbox{GeV} \ \leq \ \sqrt{s} \ \leq \ \sim \ 1.625 
\ \mbox{GeV}
\end{array}
\end{equation}

\hspace*{-2.6cm}
The minimal meromorphic parametrization is defined from the complex
pole position on the \\
\hspace*{-2.0cm}
second s - sheet, the K - plane with
$\ \Im \ K \ < \ 0 \ $ \hspace*{0.2cm} 
$\left ( \ s_{\ 0} \ = \ 4 \ m_{\ \pi}^{\ 2} \ \right )$

\vspace*{-0.0cm}
\begin{equation}
\label{eq:2}
\hspace*{-2.7cm}
\begin{array}{l}
C_{\ R}^{\ 2} \ = \ \left ( \ K_{\ R} \ - \ \frac{1}{2} \ i \ \gamma_{\ R}
\ \right )^{\ 2} \ = \ {\cal{M}}_{\ R}^{\ 2} \ - \ s_{\ 0}
\ = \ \left ( \ M_{\ R} \ - \ \frac{1}{2} \ i \ \Gamma_{\ R} \ \right )^{\ 2}
\ - \ s_{\ 0}
\vspace*{0.2cm} \\
S_{\ mmp} \ \left ( \ K_{\ R} \ , \ \gamma_{\ R} \ ; \ K \ \right ) \ = 
\ \begin{array}{c} 
\left | \ C_{\ R} \ \right |^{\ 2} \ - \ K^{\ 2} \ + \ i \ \gamma_{\ R} \ K
\vspace*{0.2cm} \\
\hline  \vspace*{-0.3cm} \\
\left | \ C_{\ R} \ \right |^{\ 2} \ - \ K^{\ 2} \ - \ i \ \gamma_{\ R} \ K
\end{array}
\end{array}
\end{equation}

\hspace*{-2.6cm}
The analytically correct derivations from solving the Roy equations in
the range limited by \\
\hspace*{-2.0cm}
Lehmann ellipses are reviewed in ref.
\cite{CCL} . The combination of scattering data , used \\
\hspace*{-2.0cm}
through absorptive parts
between $\ 0.8 \ GeV \ \leq \ M_{\ \pi \pi} \ \leq \ 2 \ GeV \ $
with {\it ideal} $\ \pi \ \pi$ scattering \\
\hspace*{-2.0cm}
lengths, accurately determined
through chiral expansions, lead to an apparently most definite \\
\hspace*{-2.0cm}
prediction and evaluation of pole parametres in the I=0 , s-wave channel 
in refs. \cite{caprini2008} 
{\it Caprini, \\ \hspace*{-2.0cm} Leutwyler} and compared 
with results obtained in ref. \cite{kampelynd} {\it Kaminski, Pelaez and
Yndurain} in \\
\hspace*{-2.0cm}
eqs. \ref{eq:3} and \ref{eq:4} below .

\hspace*{-2.6cm}
While the absolute systematic errors differ by a factor 3 - 4 , this is 
by far not a proof of the \\
\hspace*{-2.0cm}
correctness of these results, as discussed
subsequently ,
and in any case does not change the \\
\hspace*{-2.0cm}
apparent excellent agreement of
deduced phase shifts as displayed in figure 1 .

\hspace*{-2.6cm}
The evaluations following {\it Caprini} yield 4 sets
compared below with results from ref. \cite{kampelynd}

\vspace*{-0.1cm}
\begin{equation}
\label{eq:3}
\begin{array}{lll rll}
M_{\ \sigma} & = & 446 \ \pm & 6 \ \mbox{( stat )} & ^{\ + \ 40}_{\ - \ \ 4}
\ \mbox{( syst )} & \mbox{MeV}
\vspace*{0.1cm} \\
\Gamma_{\ \sigma} & = & 534 \ \pm & 12 \ \mbox{( stat )} 
& ^{\ + \ 88}_{\ - \ 66} \ \mbox{( syst )} & \mbox{MeV}
\vspace*{0.3cm} \\
M_{\ \sigma} & = & 455 \ \pm & 6 \ \mbox{( stat )} & ^{\ + \ 31}_{\ - \ 13}
\ \mbox{( syst )} & \mbox{MeV}
\vspace*{0.1cm} \\
\Gamma_{\ \sigma} & = & 556 \ \pm & 12 \ \mbox{( stat )}
& ^{\ + \ 68}_{\ - \ 86} \ \mbox{( syst )} & \mbox{MeV}
\vspace*{0.3cm} \\
M_{\ \sigma} & = & 458 \ \pm & 6 \ \mbox{( stat )} & ^{\ + \ 36}_{\ - \ 11}
\ \mbox{( syst )} & \mbox{MeV}
\vspace*{0.1cm} \\
\Gamma_{\ \sigma} & = & 506 \ \pm & 12 \ \mbox{( stat )}
& ^{\ + \ 78}_{\ - \ 56} \ \mbox{( syst )} & \mbox{MeV}
\vspace*{0.3cm} \\
M_{\ \sigma} & = & 463 \ \pm & 6 \ \mbox{( stat )} & ^{\ + \ 31}_{\ - \ 17}
\ \mbox{( syst )} & \mbox{MeV}
\vspace*{0.1cm} \\
\Gamma_{\ \sigma} & = & 518 \ \pm & 12 \ \mbox{( stat )}
& ^{\ + \ 66}_{\ - \ 68} \ \mbox{( syst )} & \mbox{MeV}
\end{array}
\end{equation}

\vspace*{-0.1cm}
\begin{equation}
\label{eq:4}
\begin{array}{lll rll}
M_{\ \sigma} & = & 496 \ \pm & 6 \ \mbox{( stat )} & \pm \ 11
\ \mbox{( syst )} & \mbox{MeV}
\vspace*{0.1cm} \\
\Gamma_{\ \sigma} & = & 516 \ \pm & 16 \ \mbox{( stat )}
& \pm \ \hspace*{0.15cm}  4 \ \mbox{( syst )} & \mbox{MeV}
\end{array}
\end{equation}
 
\rule[0.2cm]{11.5cm}{0.2mm}

\vspace*{-0.2cm}
%\hspace*{4.0cm}
\begin{center}
\hspace*{0.0cm}
\begin{figure}[htb]
% \label{fig1}
\vskip -0.6cm
\hskip -4.0cm
\includegraphics[angle=-90,width=6.5cm]{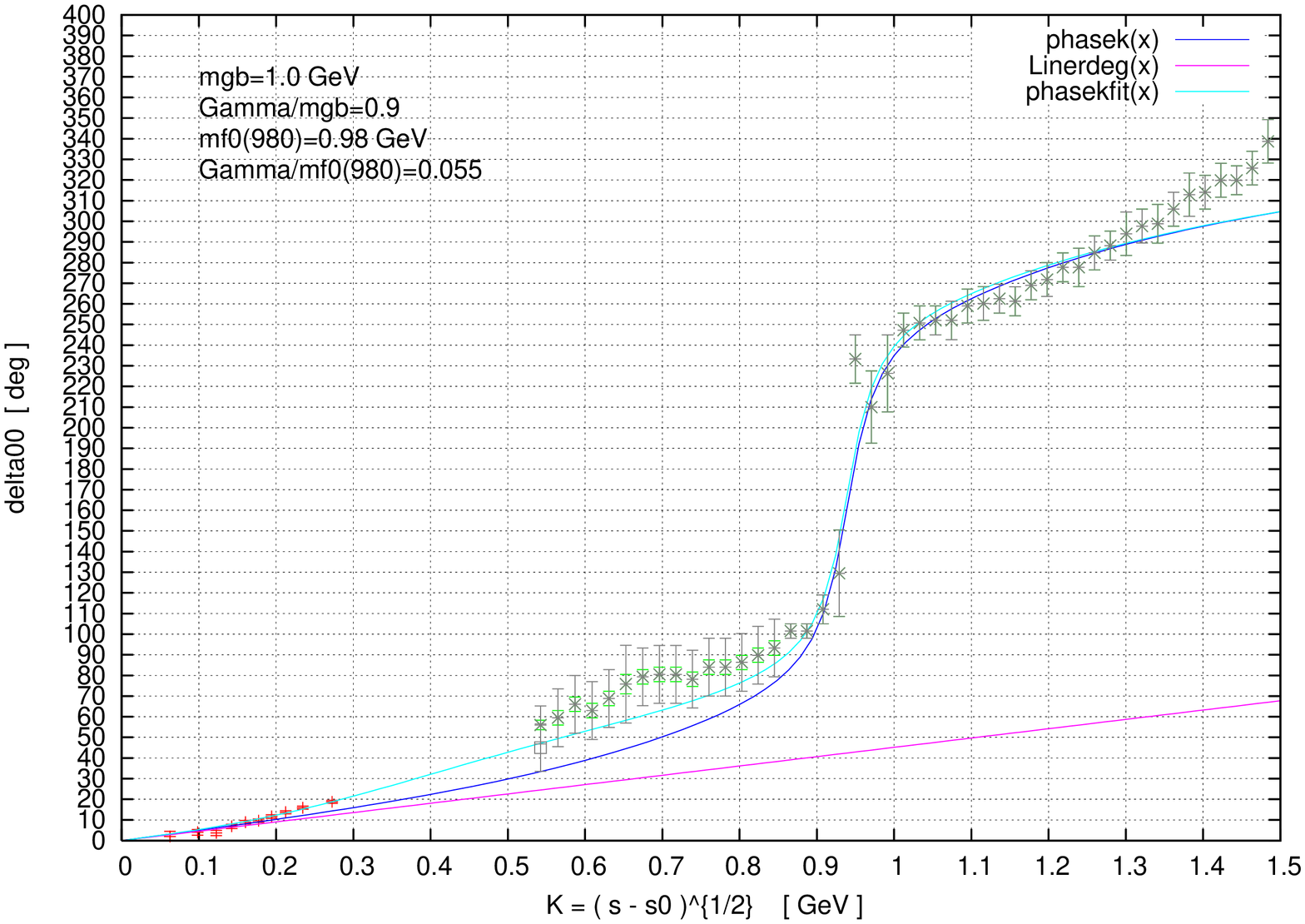}
\vskip -0.2cm
%\hskip -2.0cm
%\caption{}
%\label{fig15a}
\vspace*{+0.25cm}
{\color{red} \hspace*{-4.0cm} \vspace*{-0.15cm}
\begin{tabular}{l} Fig. 2 :
\color{red} The $\pi \pi \ , \ I=0 $ {\it ideal} elastic 
\vspace*{-0.0cm} \\
%\color{red}
s-wave
from threshold to $\ \sim \ 1.526 \ \mbox{GeV}$. 
\end{tabular}
}
\end{figure}
\vspace*{-6.0cm}
\end{center}
\hspace*{3.0cm} {\it To figure 2} : \vspace*{0.1cm} \\
\hspace*{2.7cm} \begin{tabular}[t]{l}
$M_{\ \pi \pi} \ \equiv \ \sqrt{s} \ $ throughout .
{\color{red} +} : from ref. \cite{Brigitte} ;
{\color{magenta} \rule[1.0mm]{2mm}{0.1mm} $\ \delta00 \ = \ 0.5 \ a00 \ K \ $}
\vspace*{-0.0cm} \\
as in figure 1 .
\vspace*{-0.0cm} \\
{\color{black} \rule[2.5mm]{2.1mm}{0.2mm}
\hspace*{-4.5mm}
\rule[-0.1mm]{2.1mm}{0.2mm} \hspace*{-3.2mm} \rule{0.2mm}{1.8mm}
\hspace*{-5.5mm}  \begin{tabular}{l} \vspace*{-5.5mm} \\
{\color{green} x}
\end{tabular}
%\hspace*{-0.2cm} {\color{green} x}
} \hspace*{-0.5cm} : from ref. \cite{hyams73} as in figure 1 , with enlarged
errors \\
\hspace*{0.7cm}
for systematics.
{\color{magenta} \hspace*{-0.2cm} $\rightarrow^{\ 1}$} 
\vspace*{-0.0cm} \\
{\color{black} \rule[2.5mm]{2.1mm}{0.2mm}
\hspace*{-4.5mm} \rule[-0.1mm]{2.1mm}{0.2mm}
\hspace*{-3.2mm} 
\rule{0.2mm}{2.2mm} \vspace*{-0.7mm} \hspace*{-5.5mm} 
\begin{tabular}{l} \vspace*{-0.55cm} \\
${\scriptscriptstyle \square}$ \end{tabular}
} : from ref. \cite{KamLesRyb} with statistical errors , lowest $M_{\ \pi \pi}
\ $ bin only .
\vspace*{-0.0cm} \\
\rule[0.5mm]{3.0mm}{0.1mm} \hspace*{0.8mm} : minimal meromorphic phase from the 
superposition of 
\vspace*{-0.0cm} \\
\hspace*{0.7cm} 
f0(980) and gb with masses and widths as indicated in the figure .
\vspace*{-0.0cm} \\
{\color{cyan}
\rule[0.5mm]{3.0mm}{0.1mm} \hspace*{0.8mm}
} \hspace*{-1.5mm} : background relative to the minimal meromorphic phase , 
chosen
\vspace*{-0.0cm} \\
\hspace*{0.67cm}
to follow
the lower boundary along the low fringe permitted by \\
\hspace*{0.9cm} 
{\color{black} \rule[2.5mm]{2.1mm}{0.2mm} 
\hspace*{-4.5mm}
\rule[-0.1mm]{2.1mm}{0.2mm} \hspace*{-3.2mm} \rule{0.2mm}{1.8mm}
\hspace*{-5.5mm}  \begin{tabular}{l} \vspace*{-5.5mm} \\
{\color{green} x}
\end{tabular}
%\hspace*{-0.2cm} {\color{green} x}
}
%\hspace*{-2.7mm}
%\rule[-0.1mm]{1.5mm}{0.2mm} \hspace*{-2.2mm} \rule{0.2mm}{1.5mm}
%\hspace*{-4.1mm}  \begin{tabular}{l} \vspace*{-5.5mm} \\
%{\color{green} x}
%\end{tabular}} 
{\it and} to maintain
%\vspace*{-0.0cm} \\
%\hspace*{0.60cm}
optimal agreement in the threshold- \\
\hspace*{0.7cm}
and high fringe regions. 
{\color{magenta} $\rightarrow^{\ 2}$}
\end{tabular}
\vspace*{0.2cm} 

\begin{description}
\item {\color{magenta} $\rightarrow^{\ 1}$}
The systematic error with respect to ref. \cite{hyams73} is chosen by
multiplying the quoted statistical error by the factor 2.5 , below
$ K \ = \ 0.9 \ \mbox{GeV} \ ; \ M_{\ \pi \pi} \ = \ 0.94 \ \mbox{GeV} \ $ .
This is justified here considering the difference between the 
nominal data and the minimal meromorphic phase as shown in figure 2
{\it and} from the detailed discussion of errors in ref. \cite{grayer74} . 

\item {\color{magenta} $\rightarrow^{\ 2}$} The minimal meromorphic
superposition of N resonances with identical {\it ideal} quantum numbers \\
-- in any two body channel -- corresponds to the multiplication of
the individual 
$\ S_{\ mmp} \ ( \ K_{\ R_{\ \alpha}} \ , \ \gamma_{\ R_{\ \alpha}} 
\ ; \ K \ ) \ $
factors for resonance $R_{\ \alpha} \ ; \ \alpha \ = \ 1 , \cdots, N$ as defined
in eq. \ref{eq:2} . 

\vspace*{-0.3cm}
\begin{equation}
\label{eq:5}
\begin{array}{l}
S_{\ mmp}^{\ N} \ ( \ K \ ) \ = \ \prod_{\ \alpha=1}^{\ N}
\ S_{\ mmp} \ ( \ K_{\ R_{\ \alpha}} \ , \ \gamma_{\ R_{\ \alpha}} \ ; \ K \ )
\end{array}
\end{equation}

\end{description}

\begin{description}
\item {\color{magenta} $\rightarrow^{\ 2} \ ; \ \cdots$}
The background introduced above for 
$\ J^{\ PC} \ = 0^{\ ++} \ , \ I \ = \ 0 \ ; 
\ \pi \pi \ \rightarrow \ \pi \pi \ $ is defined {\it relative} to 
$\ S_{\ mmp}^{\ N} \ $ given in eq. \ref{eq:5}

\vspace*{-0.3cm}
\begin{equation}
\label{eq:6}
\begin{array}{l}
S \ = \ S_{\ bg}^{\ N} \ S_{\ mmp}^{\ N}
\hspace*{0.2cm} ; \hspace*{0.2cm}
S_{\ bg}^{\ N} \ ( \ K \ ) \ = \ \eta_{\ bg}^{\ N} \ ( \ K \ ) \ \exp \left ( 
\ 2 \ i \ \delta_{\ bg}^{\ N} \ ( \ K \ ) \ \right )
\end{array}
\end{equation}

It follows from the meromorphic structure of $\ S_{\ mmp}^{\ N} \ $ , that
the presence in $\ T \ = \frac{1}{2 i} \ ( \ S \ - \ 1 ) \ $ of an Adler 0 , for
$\ - \ K^{\ 2} \ = \ \kappa^{\ 2} \ = \ 4 \ m_{\ \pi}^{\ 2} - s \ > \ 0 \ ;
\ \kappa \ > \ 0 \ $ requires a nontrivial \\
background $\ \rightarrow \ S_{\ bg}^{\ N} \ \not \equiv \ 1 \ $ .
For \hspace*{0.1cm} {\color{cyan} \rule[0.5mm]{3.0mm}{0.1mm} \hspace*{0.8mm}}
we use the parametrization

\vspace*{-0.3cm}
\begin{equation}
\label{eq:7}
\begin{array}{l}
\delta_{\ bg}^{\ 2} \ ( \ K \ ) \ = \ ( \ K \ / \ K_{\ 1} \ )^{\ 3} 
\ e^{\ - \ B \ K^{\ 2}} 
\hspace*{0.2cm} ; \hspace*{0.2cm} 
K_{\ 1} \ = \ 0.59 \ \mbox{GeV} \ , \  B \ = \ 4.2 \ \mbox{GeV}^{\ -2}
\vspace*{0.2cm} \\
\eta_{\ bg}^{\ 2} \ ( \ K \ ) \ = \ 1 \ ; 
\ \mbox{with modifications particularly for} \ N=2 \ \rightarrow \ N=3
\end{array}
\end{equation}

concentrating on the low fringe region , and coming back to inelasticities in
the high fringe below in conjunction with the third resonance f0(1500) and
figure 3 . I follow the hypotheses and derivations presented
in refs. \cite{PMWO} and concerning the role of f0(1500) in the decays
$\ B \ \rightarrow \ K \ \pi \pi \ , \ K \ \overline{K} K \ $
\cite{PMWOb} in collaboration with Wolfgang Ochs .
\vspace*{0.2cm}
\end{description}
%\hspace*{5.3cm} 
\vspace*{0.3cm} 

\hspace*{5.3cm} {\it To figure 3} : \\
\hspace*{5.3cm} \begin{tabular}[t]{l}
This is an extension of figure 2 to include \\
the influence of 
three resonances  
%\vspace*{-0.0cm} \\
f0(980) , gb and f0(1500). \\ 
{\color{red} +} , 
{\color{magenta} \rule[1.0mm]{2mm}{0.1mm}} 
\hspace*{0.40cm} : as in figure 1 .
\vspace*{-0.0cm} \\
{\color{red} \rule[2.5mm]{2.1mm}{0.2mm}
\hspace*{-4.5mm}
\rule[-0.1mm]{2.1mm}{0.2mm} \hspace*{-3.2mm} \rule{0.2mm}{1.8mm}
\hspace*{-5.5mm}  \begin{tabular}{l} \vspace*{-5.5mm} \\
x
\end{tabular}
%\hspace*{-0.2cm} {\color{green} x}
} ,
%\rule[1.5mm]{1.5mm}{0.2mm}
%\hspace*{-2.7mm}
%\rule[-0.1mm]{1.5mm}{0.2mm} \hspace*{-2.2mm} \rule{0.2mm}{1.5mm}
%\hspace*{-4.1mm}  \begin{tabular}{l} \vspace*{-5.5mm} \\
%x
%\end{tabular}
%} 
 \hspace*{0.0cm} {\color{red} \rule[2.5mm]{2.1mm}{0.2mm}
\hspace*{-4.5mm} \rule[-0.1mm]{2.1mm}{0.2mm}
\hspace*{-3.2mm}
\rule{0.2mm}{1.8mm} \vspace*{-0.7mm} \hspace*{-5.5mm}
\begin{tabular}{l} \vspace*{-0.6cm} \\
${\scriptscriptstyle \square}$ \end{tabular}
}
\hspace*{-3.4mm} : as in figure 2 except for the color .
\vspace*{-0.0cm} \\
\rule[0.5mm]{3.0mm}{0.1mm} \hspace*{0.9cm} : minimal meromorphic phase from the 
superposition of gb 
\vspace*{-0.0cm} \\
\hspace*{1.4cm}  and f0(980) but with different f0 mass 
$\ mf0 \ = \ 0.99 \ \mbox{GeV} \ $ , 
\vspace*{-0.0cm} \\
\hspace*{1.40cm} same ratio $\Gamma_{\ f0} \ / \ mf0 \ = \ 0.055$ .
\vspace*{-0.0cm} \\
{\color{cyan} \rule[0.5mm]{3.0mm}{0.1mm}} 
\hspace*{0.9cm} : background phase added with same mass and width
\vspace*{-0.0cm} \\
\hspace*{1.40cm}
parameters as for \hspace*{0.1cm} \rule[0.5mm]{3.0mm}{0.1mm} \hspace*{0.1cm}
and $K_{\ 1} \ = \ 0.62 \ \mbox{GeV}$ ( eq. \ref{eq:7} ) .
\vspace*{-0.0cm} \\
{\color{cyan} \rule[0.5mm]{0.5mm}{0.5mm} \hspace*{0.3mm}
\rule[0.5mm]{0.5mm}{0.5mm} \hspace*{0.3mm} \rule[0.5mm]{0.5mm}{0.5mm}}
\hspace*{0.7cm} : as 
\hspace*{0.1cm} {\color{cyan} \rule[0.5mm]{3.0mm}{0.1mm}} \hspace*{0.1cm} 
in figure 2 , 
\vspace*{-0.0cm} \\
\hspace*{1.4cm}
with 
$\ mf0 \ = \ 0.98 \ \mbox{GeV} \ $ , $\Gamma_{\ f0} \ / \ mf0 \ = \ 0.055$ .
\vspace*{-0.0cm} \\
{\color{black} \rule[0.5mm]{3.0mm}{0.1mm}}
\hspace*{0.9cm} :  minimal meromorphic phase from the superposition of
\vspace*{-0.0cm} \\
\hspace*{1.4cm} gb , f0(980) {\it and} f0(1500) with mass and width
parameters
\vspace*{-0.0cm} \\
\hspace*{1.4cm} $mf0(1500) \ = \ 1.51 \ \mbox{GeV} \ $ , 
$\Gamma_{\ f0(1500)} \ / \ mf0(1500) \ = \ 0.07$ ,
\vspace*{-0.0cm} \\
\hspace*{1.4cm} and background parameters $\ \eta_{\ bg}^{\ 3} \ = \ 1 \ $
\vspace*{-0.0cm} \\
\hspace*{1.4cm}
to keep qualitative features  
%\vspace*{-0.0cm} \\
of f0(1500) only 
\vspace*{-0.0cm} \\
\hspace*{1.4cm}
and $\ K_{\ 1} \ = \ 0.62 \ \mbox{GeV} \ , \ B \ = \ 4.2 \ \mbox{GeV}^{\ -2}$ 
( eq. \ref{eq:7} ) .
\vspace*{-0.0cm} \\
\hspace*{1.4cm} The rise of the s-wave phase towards the end of the 
\vspace*{-0.0cm} \\
\hspace*{1.4cm}  high fringe region was remarked
in ref. \cite{kampelynd} .
\vspace*{-0.0cm} \\
\hspace*{1.4cm}
It formed the entry point of
the discussion in ref. \cite{PMQCD2008} .
\end{tabular}

\vspace*{-7.5cm}
%\hspace*{4.0cm}
\begin{center}
\hspace*{0.0cm}
\begin{figure}[htb]
% \label{fig1}
\vskip -2.1cm
\hskip -1.6cm
\includegraphics[angle=0,width=6.0cm]{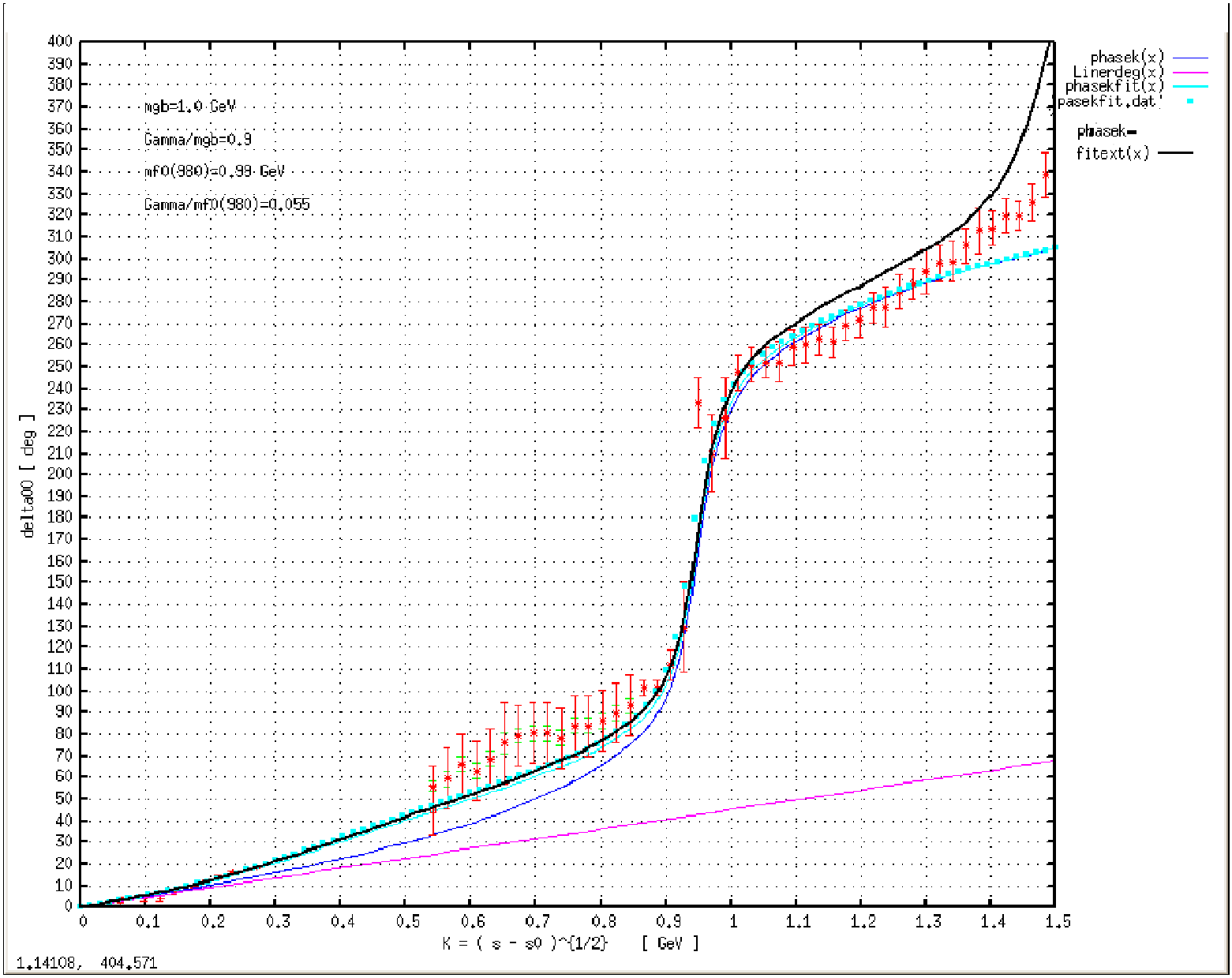}
\vskip -0.0cm
%\hskip -2.0cm
%\caption{}
%\label{fig15a}
\vspace*{+0.20cm}
{\color{red} \hspace*{-2.0cm} \vspace*{-0.15cm}
\begin{tabular}{l} Fig. 3 :
\color{red} The $\pi \pi \ , \ I=0 $ {\it ideal} elastic \\
s-wave  
% \vspace*{-0.1cm} \\
%\color{red}
from threshold to $\ \sim \ 1.526 \ \mbox{GeV}$. 
\end{tabular}
}
\end{figure}
\end{center}

\newpage

\vspace*{-0.5cm}
%\hspace*{4.0cm}
\begin{center}
\hspace*{0.0cm}
\begin{figure}[htb]
% \label{fig1}
\vskip -1.4cm
\hskip -3.4cm
\includegraphics[angle=0,width=6.0cm]{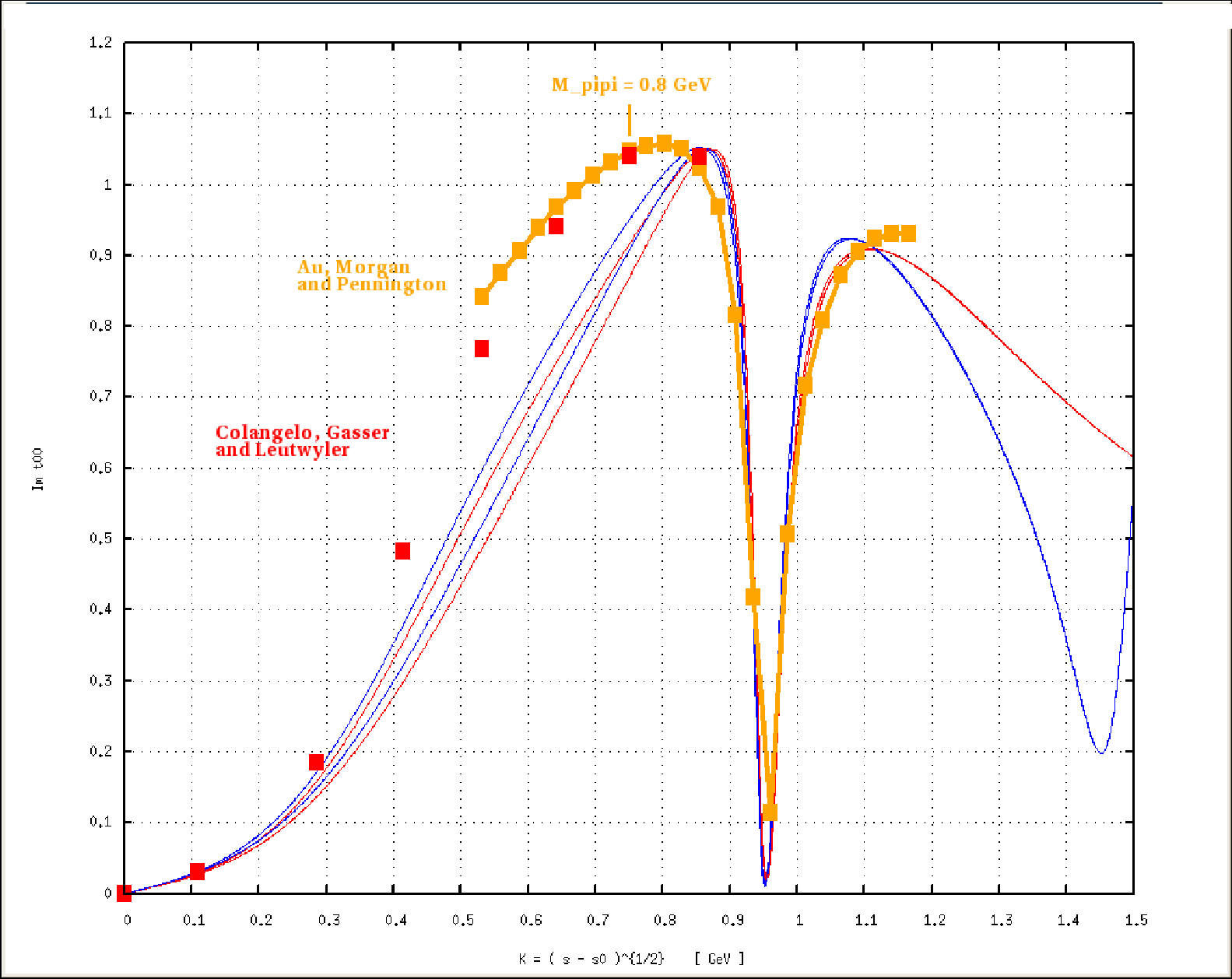}
\vskip -0.0cm
%\hskip -2.0cm
%\caption{}
%\label{fig15a}
\vspace*{+0.20cm}
{\color{red} \hspace*{-3.6cm} \vspace*{-0.00cm}
\begin{tabular}{l} Fig. 4 :
\color{red} The imaginary part of the \\
$\pi \pi \ , \ I=0 $ {\it ideal} 
elastic s-wave  
%\vspace*{-0.0cm} \\
%\color{red}
from threshold to $\ \sim \ 1.526 \ \mbox{GeV}$. 
\end{tabular}
}
\end{figure}
\end{center}

\vspace*{-6.8cm}
\hspace*{2.5cm}
\begin{tabular}[t]{l}
{\it To figure 4}
\end{tabular}
: \\
\hspace*{3.1cm} \begin{tabular}[t]{l}
Here we present aspects of the absorptive part $\ \Im \ t_{\ 00} \ $
\vspace*{-0.0cm} \\
with
$\ t_{\ 00} \ = \ \left ( \ M_{\ \pi \pi} \ / \ K \ \right ) \ \frac{1}{2 i}
\ \left ( \ S_{\ bg}^{\ N} \ S_{\ mmp}^{\ N} \ - \ 1 \ \right ) 
\ ; \ N \ = \ 2,3 \ $ 
\vspace*{0.0cm} \\
and compare with the analyses of \cite{AuMoPenn} 
{\it Au, Morgan and Pennington},  
\vspace*{-0.0cm} \\
\cite{CGL} {\it Colangelo, Gasser and Leutwyler} and the solution
to the 
\vspace*{-0.0cm} \\
Roy equations \cite{ACGL}
{\it Ananthanarayan, Colangelo, Gasser and Leutwyler}.
\vspace*{-0.0cm} \\
The resonance parameters used for $\ N=2 \ $ and $\ N=3 \ $ are
\vspace*{-0.0cm} \\
\begin{tabular}{lll lll l}
$mf0$ & = & 0.99 & GeV & 
$\Gamma_{\ f0} \ / \ mf0$ & = & 0.055
\vspace*{-0.0cm} \\
$mgb$ & = & 1.0 & GeV & $\Gamma_{\ gb} \ / \ mgb$ & = & 0.9
\vspace*{-0.0cm} \\
$mf0(1500)$ & = & 1.51 & GeV & 
$\Gamma_{\ f0(1500)} \ / \ mf0(1500)$ & = & 0.07 
\end{tabular}
\vspace*{1.7cm}
\end{tabular}

\hspace*{-5,0cm} \begin{tabular}{l}
The inelasticity is extended to include ( for $\ \pi \pi \ $ elastic )
two $I \ = \ 0$ 
$\pi \pi$ and $K \overline{K}$ two-body channels
\end{tabular}
\vspace*{-0.3cm}

\hspace*{-3.0cm}
\begin{equation}
\label{eq:8}
\vspace*{-0.4cm}
\begin{array}{l}
\eta_{\ bg}^{\ 2,3} \ ( \ K \ ) \ =
\ \vartheta \ ( K_{\ th} \ - \ K \ ) \ +
\ \vartheta \ ( K \ - \ K_{\ th} \ ) \ a \ e^{\ - \ b \ K^{\ '} \ / 
\ K^{\ '}_{\ 15}} 
\vspace*{0.2cm} \\
K^{\ '} \ ( \ K \ ) \ = \ \left ( \ K^{\ 2} \ - \ K_{\ th}^{\ 2} 
\ \right )^{\ 1/2}
\hspace*{0.2cm} ; \hspace*{0.2cm}
K_{\ th} \ = \ \left ( \ 4 \ m_{\ K}^{\ 2} \ - \ 4 \ m_{\ \pi}^{\ 2}
\ \right )^{\ 1/2}
\vspace*{0.2cm} \\
K^{\ '}_{\ 15} \ = \ \left ( \ mf0(1500)^{\ 2} \ - \ 4 \ m_{\ K}^{\ 2}
\ \right )^{\ 1/2}
\hspace*{0.2cm} ; \hspace*{0.2cm}
m_{\ K} \ = \ 0.49565 \ \mbox{GeV}
\vspace*{0.2cm} \\
\mbox{with parameters fixed at}
\hspace*{0.3cm} 
a \ = \ 1 
\hspace*{0.2cm} ; \hspace*{0.2cm}
b \ = \ - \ \log \ 0.6 \ = 0.5108
\hspace*{0.2cm} ; \hspace*{0.2cm}
m_{\ \pi} \ = \ 0.13957 \ \mbox{GeV}
%eta(K)=0.5*(1.-sign(K-Kth))+0.5*(1.+sign(K-Kth))*Aeta15*beta(K)
\end{array}
\end{equation}
\vspace*{-0.2cm}

\hspace*{-3.0cm} \begin{tabular}{l}
No data is used to determine the elasticity parameter 
-- $\eta_{\ bg}^{\ 2,3} \ ( \ M_{\ \pi \pi} \ = \ mf0(1500) \ ) \ \sim
\ 0.6$. 
\end{tabular}

\hspace*{-3.0cm}
\vspace*{-0.0cm}
\begin{tabular}[t]{l}
{\it To figure 4}
\vspace*{-0.0cm} \\
{\it (continued)}
\end{tabular}
: \begin{tabular}[t]{l}
{\color{red} \rule[0.5mm]{3.0mm}{0.1mm}}
\hspace*{0.9cm} : from $\ S \ = \ S_{\ bg}^{\ 2,3} \ S_{\ mmp}^{\ 2} \ $
\vspace*{-0.0cm} \\
\hspace*{1.4cm} with parameters as specified above 
( and eqs. \ref{eq:7} , \ref{eq:8} ) as in figure 3 
\vspace*{-0.0cm} \\
\hspace*{1.4cm}
except : 
\begin{tabular}[t]{l}
upper curve $\ \rightarrow \ K_{\ 1} \ = \ 0.59 \ \mbox{GeV} \ $ as in figure 2
\vspace*{-0.0cm} \\
lower curve $\ \rightarrow \ K_{\ 1} \ = \ 0.67 \ \mbox{GeV} \ $ .
\end{tabular}
\vspace*{-0.0cm} \\
\rule[0.5mm]{3.0mm}{0.1mm}
\hspace*{0.9cm} : from $\ S \ = \ S_{\ bg}^{\ 2,3} \ S_{\ mmp}^{\ 3} \ $
\vspace*{-0.0cm} \\
\hspace*{1.4cm} with parameters as specified above
( and eqs. \ref{eq:7} , \ref{eq:8} ) as in figure 3
\vspace*{-0.0cm} \\
\hspace*{1.4cm} except :
\begin{tabular}[t]{l}
upper curve $\ \rightarrow \ K_{\ 1} \ = \ 0.59 \ \mbox{GeV} \ $ as in figure 2
\vspace*{-0.0cm} \\
lower curve $\ \rightarrow \ K_{\ 1} \ = \ 0.67 \ \mbox{GeV} \ $ .
\end{tabular}
\vspace*{-0.0cm} \\
{\color{brown} \rule[0.5mm]{3.0mm}{0.1mm} \hspace*{-0.65cm}
\rule[-0.3mm]{1.5mm}{1.5mm} \hspace*{0.05cm} \rule[-0.3mm]{1.5mm}{1.5mm}}
\hspace*{0.9cm} : $\ \Im \ t_{\ 00} \ $ from ref. \cite{AuMoPenn} .
\vspace*{-0.0cm} \\
\hspace*{-0.25cm} {\color{red} \rule[-0.3mm]{1.5mm}{1.5mm} \hspace*{0.05cm}
\rule[-0.3mm]{1.5mm}{1.5mm}}
\hspace*{0.9cm} : $\ \Im \ t_{\ 00} \ $ from ref. \cite{CGL} .
\end{tabular}

\rule{11.5cm}{0.3mm}

Concluding remarks
\vspace*{-0.0cm}

\begin{description}
\item 1) the main analyses of reactions 
$ \ \pi {\cal{N}} \ \rightarrow \ \pi \pi {\cal{N}} 
\ \left ( \ \Delta \ \right ) \ $ in refs. \cite{Protopop73} , \cite{hyams73}
cannot be taken at face value for the derived elastic $\ \pi \pi\ $ 
s-waves within the quoted errors , in both low and high
fringe regions \\
( defined in eq. \ref{eq:1} ) ,
\vspace*{-0.1cm}

\item 2) derivations and hypotheses discussed in refs. \cite{PMWO} ,
\cite{PMWOb} are basically correct ,
\vspace*{-0.1cm}

\item 3) claims of a scalar resonance pole in the region
within a radius of at least 150 MeV around the position
$\ \sqrt{s} \ = \ 500 \ - \frac{i}{2} \ 500 \ \mbox{MeV} \ $ on the 
second s-sheet of elastic
$\ \pi \pi \ $ scattering are incorrect .
\vspace*{-0.1cm}
\end{description}

I wish to dedicate this work to the memories of Jan Stern, Francisco \\
\hspace*{0.4cm} Yndurain and Peter Schlein .
}

\newpage

{\color{blue}

}

\end{document}